# *In-Situ Analysis of Vibration and Acoustic Data in Additive Manufacturing*

Muhammad Fasih Waheed
*Electrical and Computer Engineering*
*Florida A&M University*
Tallahassee, USA
muhammad1.waheed@famu.edu

Dr. Shonda Bernadin
*Electrical and Computer Engineering*
*Florida A&M University*
Tallahassee, USA
bernadin@eng.famu.fsu.edu

***Abstract*— Vibration from an erroneous disturbance harms the manufactured components and lowers a FDM printer's output quality. For moving machinery, vibration analysis and control are crucial. The additive manufacturing technique is the basis for 3D printing which utilizes mechanical movements of extruder to print objects and faults occur due to unwanted vibrations. Therefore, it is vital to examine the 3D printer's vibration pattern. In this work we will observe these parameters of FDM printer exemplified by MakerBot method X. To analyze, it is necessary to understand the motion it generates, and appropriate sensors must be selected to detect those motions. The sensor measurement values can be used to determine the printer's condition. We have used an accelerometer and an acoustic sensor to measure the vibration and acoustics produced by the printer. The individual output from these sensors are examined. The findings show that vibration occurs at relatively low levels during continuous motion because it only occurs at the components' soft edges. Due to the abrupt acceleration and deceleration during the zigzag motion, the vibration is at its peak.**

## 1. Introduction

To ensure the quality and consistency of parts produced by the MakerBot Method X, it is essential to have an effective quality control system in place. One of the most promising approaches for quality control in FDM is condition monitoring. Condition monitoring involves the continuous monitoring of key parameters during the printing process to detect anomalies and identify potential defects before they become critical. Many studies have already been done on this topic [1-8] This technique can provide valuable insights into the quality of printed parts, enabling manufacturers to take corrective action and optimize their production processes.

In this paper, we will review the current state-of-the-art in-situ condition monitoring for FDM printers, examine the benefits and challenges associated with this technique, propose future research directions to enhance the effectiveness of condition monitoring in this printer, and present experimental findings to contribute valuable insights to the field.

FDM is a prevalent additive manufacturing method that involves melting material, extruding it through a nozzle, depositing it on a bed, and layering it to form parts. FDM offers several benefits, including low cost, simple operation, and the ability to manufacture complex shapes efficiently and environmentally friendly. We are going to differentiate between various modes of failure within the 3d printer caused by the motions. We will also analyze the acoustic patterns of the 3d printer during motion.

### *A. Background*

Despite technological advancements, FDM printing can still result in defects, necessitating extensive research into monitoring the printing process and product quality to enhance success rates and efficiency. [1] By doing a comparison and analysis of vibration signal generated by extruder head movements in a FDM printer we can further understand the root cause of failure, in FDM printer the x-axis, y-axis, and z-axis motors work together to control the movement of the extruder head and the build platform in three-dimensional space.

The print head moves in 2 axis 'x' and 'y', The print bed moves in 'z' direction, The x-axis motor is responsible for moving the print head horizontally along the x-axis, while the y-axis motor controls the movement of the build platform along the y-axis. The z-axis motor, on the other hand, controls the vertical movement of the print head and the build platform along the z-axis. Also three different motions, including point-to-point, zigzag, and continuous motion are produced by extruder which result in sometimes unwanted vibration. Fig 1.1. shows the location of X and Y axis motor in the 3D printer.

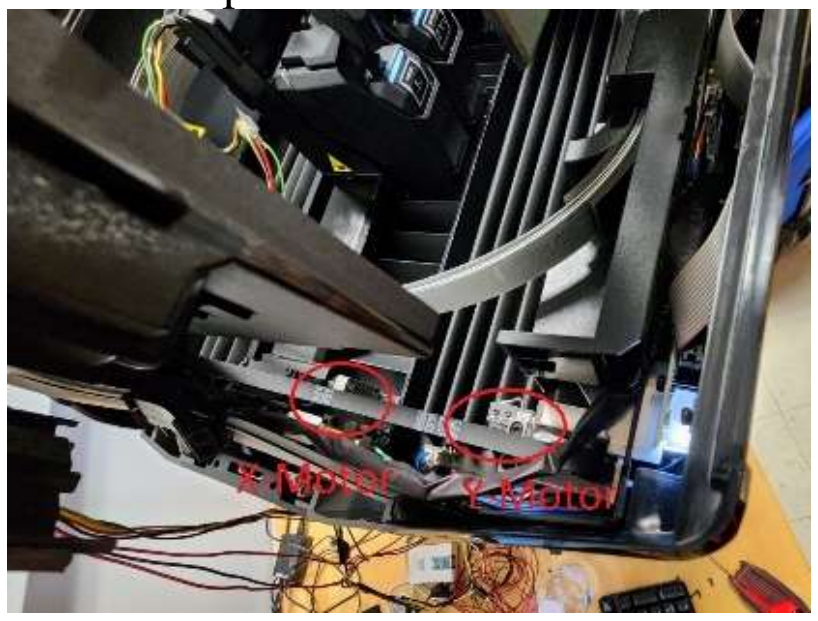

Fig. 1.1. Top view shows highlighted X and Y axis motor.

For this research, we will be using The MakerBot Method X, it is a high-performance 3D printer that uses Fused Deposition Modeling (FDM) technology to produce high-quality parts. This printer is designed for industrial applications and is capable of printing with a wide range of materials, including ABS, Nylon, and Carbon Fiber.

## II. IN- SITU MONITORING AND ANALYSIS OF 3D PRINTER HEALTH

In the context of 3D printing, in situ monitoring (ISM) refers to the use of automated processes to facilitate real-time quality monitoring during the printing process. This type of monitoring is essential for detecting product quality defects and ensuring the overall health of the 3D printing process [9][10]. Research and studies have explored the use of accelerometers and acoustic sensors for in situ monitoring of 3D printers, particularly in the context of fused filament fabrication (FFF) processes. Accelerometers have been utilized to analyze and control the vibration of 3D printers, with the extracted acceleration data being used to identify various states of the FFF machine and predict product quality[14][15]. Additionally, the installation of accelerometers on 3D printers has been presented in experimental setups for nozzle condition monitoring, demonstrating their potential in this application[16].

Acoustic sensors have also been considered for in situ monitoring during the FFF process. Some studies have used time-domain features of acoustic emission data to identify abnormal conditions, such as material runout and blocked extruders, showcasing the potential of acoustic sensors in monitoring the 3D printing process [16]. Various techniques, such as in-situ print characterization, defect monitoring via conductive filament and Ohm's Law, and digital image correlation (DIC)-based monitoring methods, have been developed for in situ monitoring of additive manufacturing [11][12]. ASTM International has released a report on in-situ monitoring for 3D printing, which aims to provide a comprehensive assessment of the technology readiness and to identify gaps and challenges within the field of in-situ technology [19]. The report is the result of a workshop series organized by ASTM International, NASA, and America Makes, and it serves as a valuable resource for the 3D printing community [10][13].

### A. *Components in a 3D printer*

A general idea of main components in a 3D printer are shown in Fig 1.2. [17]. The thermoplastic filament is driven from the extruder cold end to the hot end, where it is heated until molten, then fed through the nozzle to be bonded layer-by-layer until the product is complete. In some cases, the platform may occasionally be referred to as just a build platform and not heated. Machine downtime and material loss can be prevented by keeping an eye on a 3D printer's health while it is printing. The extruder and heated build platform are the two crucial components that, to a significant extent, define the success of printing for the majority of FFF machines used by people and industry [9]. Nozzle clogging, nozzle temperature, filament runout, and hot build are a few instances of specific situations or factors that need to be watched.

The cold end and the hot end are the two components that make up the extruder. Fig 1.2. shows the specifics for these two parts. Two crucial parts, the idler pulley shaft, which holds the filament, the stepper motor and drive gear make up the cold end, which pushes the filament into the hot end of the extruder: [10]. The heater block, heat sink, heater cartridge, and thermocouple in the hot end work together to precisely heat the filament to a liquid condition before extruding it from the nozzle when the filament reaches the hot end [11].

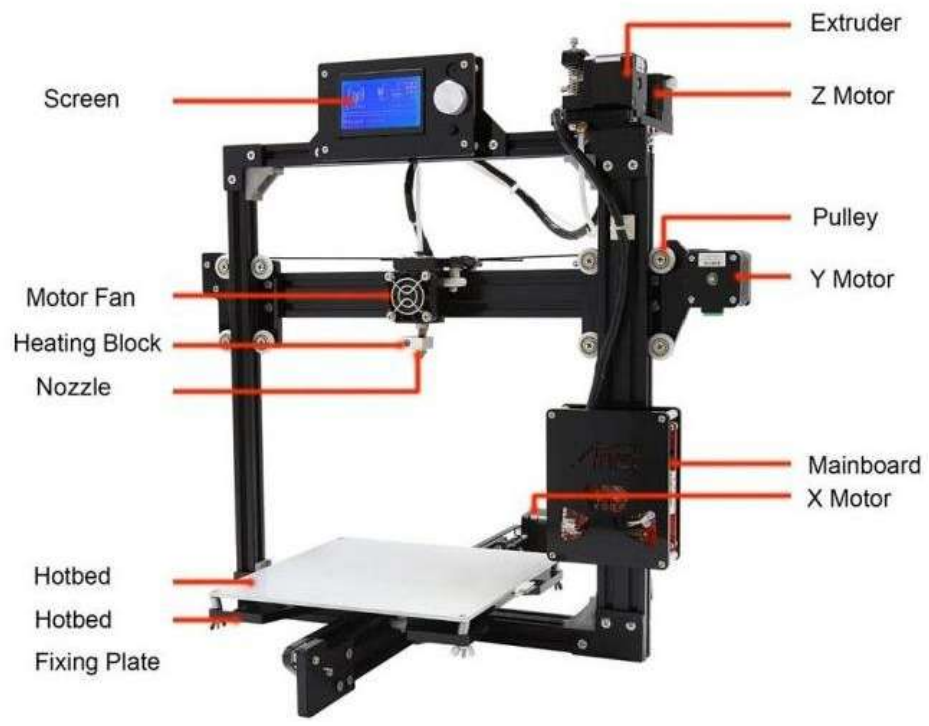

Fig. 1.2. Shows several components of an FDM printer.

#### 1) *Extruder teardown*

To gain a deeper understanding of the intricate mechanics and electrical systems at play within a print extruder, a model 1a extruder is disassembled. This dissection allowed for a comprehensive exploration of the inner workings of this critical component in 3D printing technology. One of the key revelations during this disassembly is the identification of two primary controllers integral to the extruder's operation and communication with the 3D printer. These controllers played distinct but complementary roles in ensuring the accurate and efficient extrusion of material. The first of these controllers, denoted as GQ255, is primarily responsible for overseeing and regulating the extrusion process. See Fig 1.3. its key function is to receive signals emanating from a magnetic angle sensor, which had been meticulously affixed to the filament feeding tube.

The magnetic angle sensor, through its continuous monitoring of the position and rotation of the filament feeding tube, provided precise data that was transmitted to controller GQ255. It uses real-time data to adjust filament speed and flow for precise and aligned 3D printing. The magnetic angle sensor's synchronization with GQ255 is crucial for precision. The second controller manages communication between print extruder and 3D printer using serial communication. It enables real-time adjustments and monitoring of parameters like temperature and print speed for quality prints.

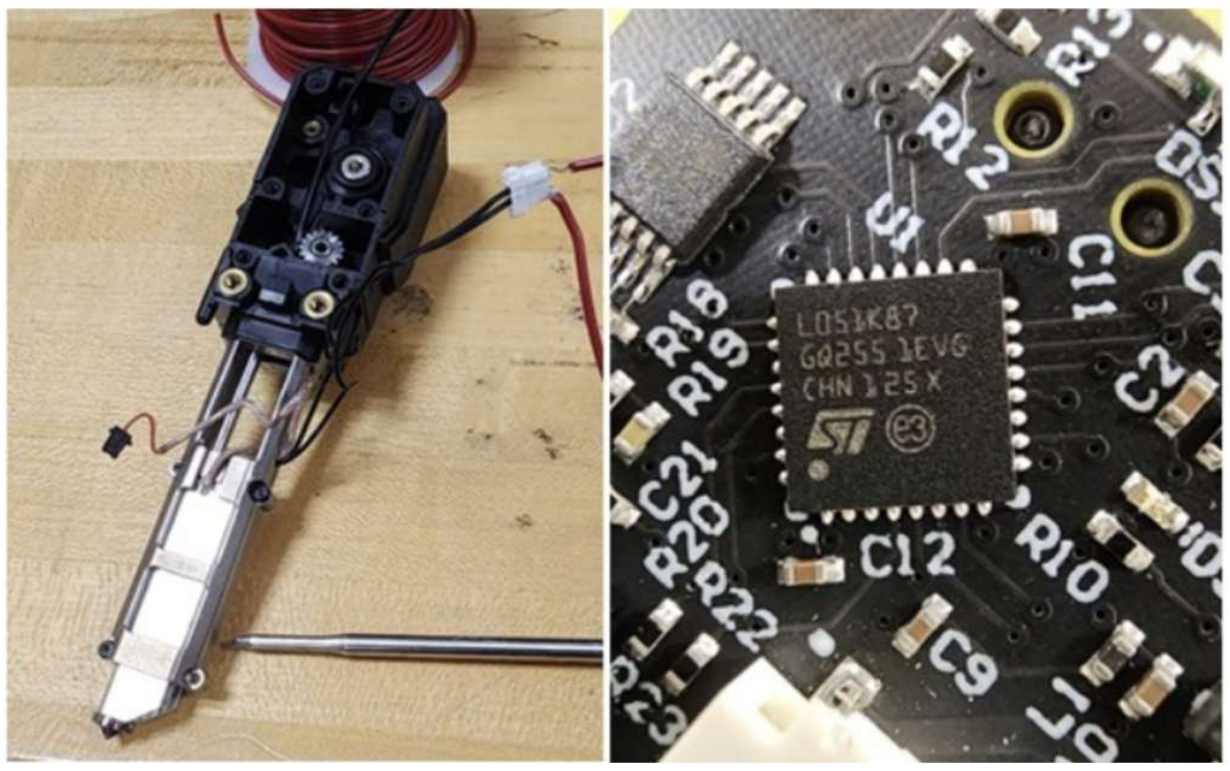

Fig. 1.3. Mechanical and electronic components of model 1A extruder.

*B. Vibration signal preprocessing*

To achieve desirable outcomes while analyzing vibration signals, it is essential to consider multiple factors in practicality. The first factor is the signal's quality, which holds significant importance. As per research [12], more than 70% of success is attributed to good signal acquisition. To enhance the signal quality and guarantee the consistency of the results, we are using ADXL 335 accelerometer from Adafruit industries as a vibration sensor, See Fig 1.4. To digitize the vibration signal accurately, it is essential to determine the appropriate sampling rate. The Nyquist theorem states that the sampling frequency must be a minimum of two times higher than the maximum frequency in the vibration signal to capture all significant information. However, an excessively high sampling rate can result in increased computational costs. Therefore, in the experiments, a sampling rate of 2KHz has been employed to strike a balance between preserving information integrity and ensuring computational efficiency.

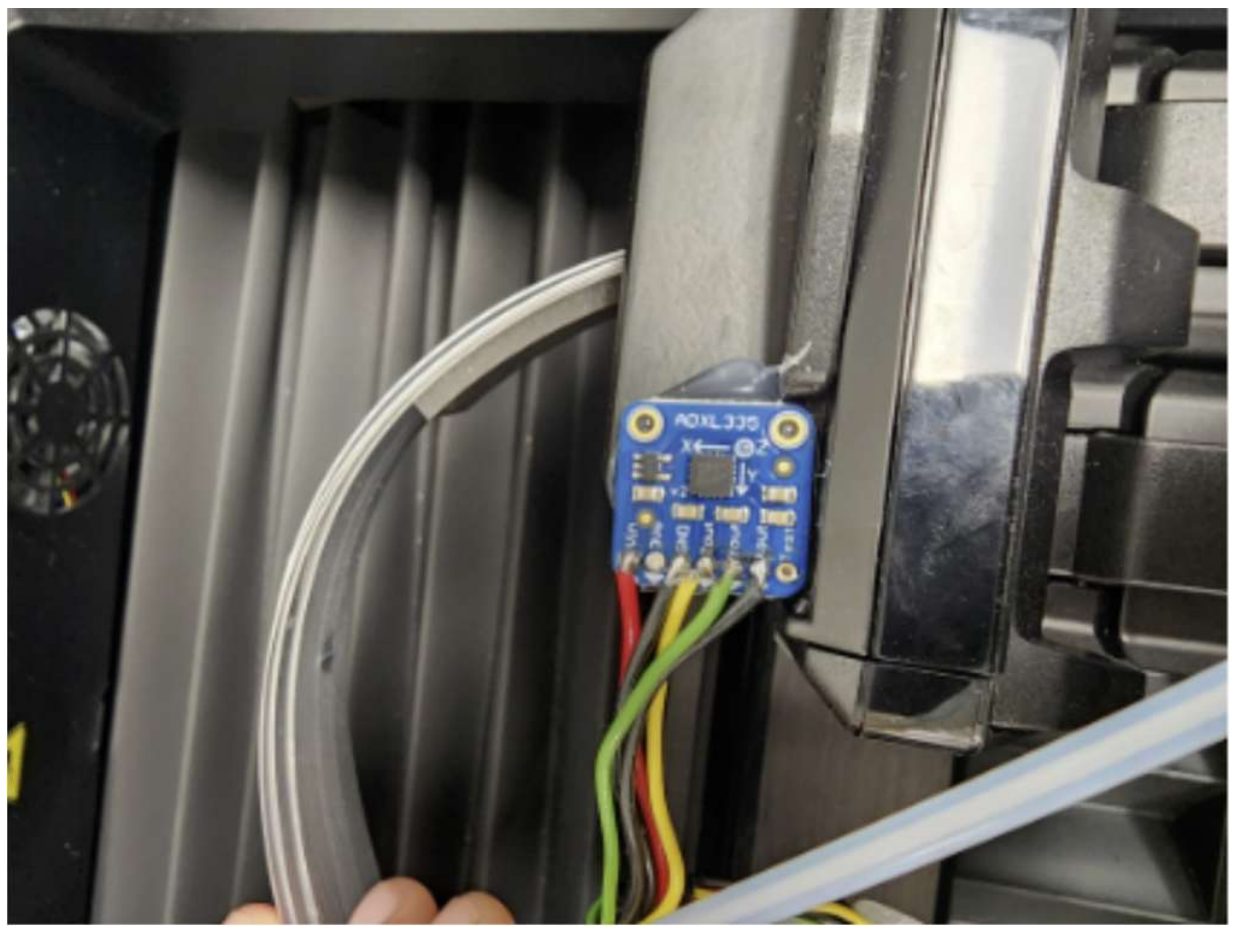

Fig. 1.4. Accelerometer installed on top of the extruder carriage.

To differentiate operational states while FFF printing, the extracted characteristics must possess a high sensitivity to operating conditions as opposed to being impervious to noise. This study [13] has presented five frequently employed features in time-domain signals, namely mean, STD, RMS, CF, and KI. For a vibration signal x(n), where N, u, and $\sigma$ represent length, mean, and STD values, respectively, RMS, CF, and KI can be formulated as follows. The root mean square (RMS) is directly proportional to the energy levels of the signal in the temporal domain. Variations in the RMS could indicate alterations in the operational states of the three-dimensional printer or be linked to defects in the final product [13].

$$RMS = \frac{1}{N}\sum_{i=1}^{N}(x(i))^2$$

CF is the ratio of peak-to-valley value to the RMS value of the vibration signal and elucidates any outcome present in the vibration signal [14].

$$CF = \frac{\max\big(x(n)\big) - \min\big(x(n)\big)}{RMS}$$

The implementation of ZigZag motion in 3D printers is recognized for its efficacy in achieving intricate prints; however, it necessitates a careful consideration of the accompanying vibrational effects. ZigZag motion inherently generates substantial vibrations during the rapid and repeated directional changes of the print head. These vibrations, characterized by the introduction of harmonic frequencies, have the potential to adversely impact various mechanical components within the printer. Of particular concern is the likelihood of compromised structural integrity due to the vibrations, which may manifest as consequences such as the loosening of screws or vulnerabilities in the design of metal sheets. These structural deficiencies, if unaddressed, can escalate to significant issues, posing a threat to the overall stability of the printer. Additionally, the vibrational effects of Zig Zag motion extend to critical mechanical parts, with pulleys and drive gears emerging as susceptible elements.

The oscillations induced by Zig Zag motion may result in misalignment, slippage, or other operational irregularities in these components, thereby necessitating a comprehensive understanding and proactive mitigation strategies to ensure the sustained performance and durability of 3D printers.

*C. Acoustic signal preprocessing*

Audio sensors can detect the high and low frequency acoustic emissions generated by the movement of the motors, gears and the vibrations they create [8]. These sensors can detect any changes in the acoustic emissions, such as increased noise or unusual sounds, which can indicate a problem with the system. Fig 1.5. shows an illustration of how audio sensor can be used to visualize frequency patterns.

Multiple arrays of acoustic sensors can be placed on a 3D printer to enhance mechanical fault detection. The sensors can be placed at strategic locations on the printer, such as near the motors, gears, and belts. The sensors can be used to detect acoustic emissions, which are sound waves that are generated by the printer during operation. These sensors are part of a larger system that includes various other types of sensors, such as accelerometers, vibration sensors, thermal cameras, and optical cameras, to monitor the health of the printer during the printing process [18]. The use of multiple sensors allows for a more comprehensive understanding of the printer's performance, as different sensors can capture different aspects of its operation [19].

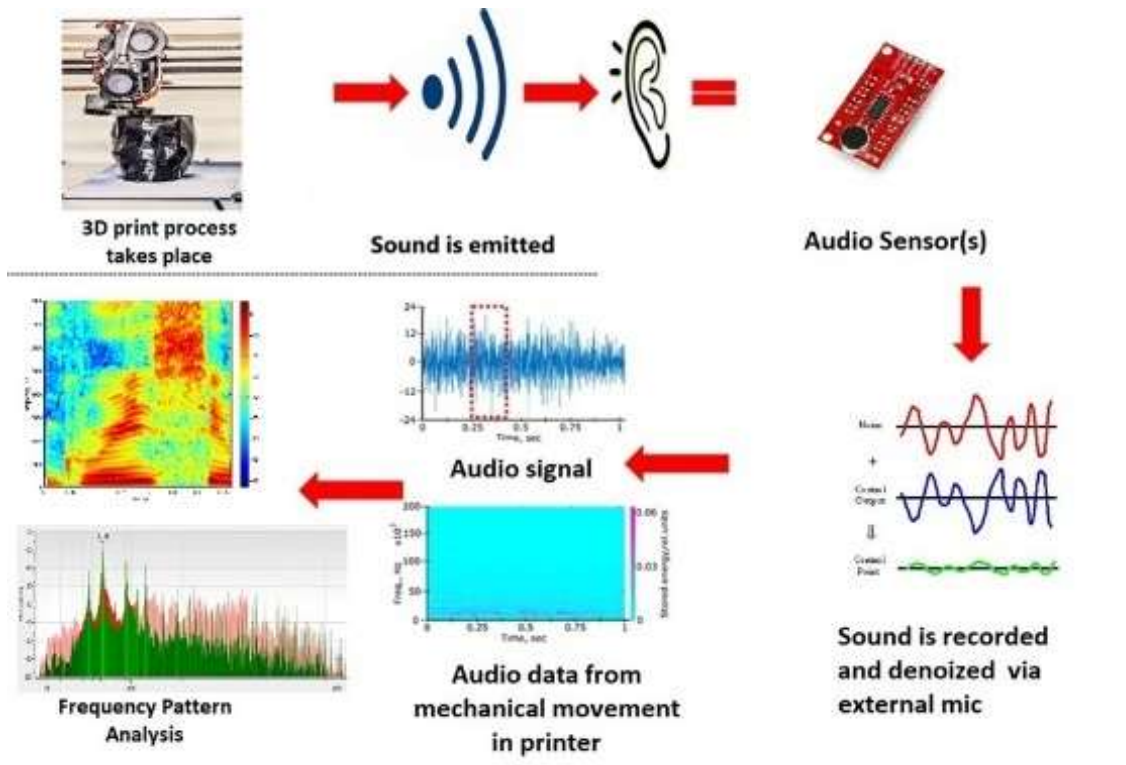

Fig. 1.5. Illustration of utilization of audio sensor for frequency pattern visualization and analysis.

## III. EXPERIMENTAL SETUP

This aspect of the study's experimental design is described in detail, and Fig 2.1. shows the associated flowchart.

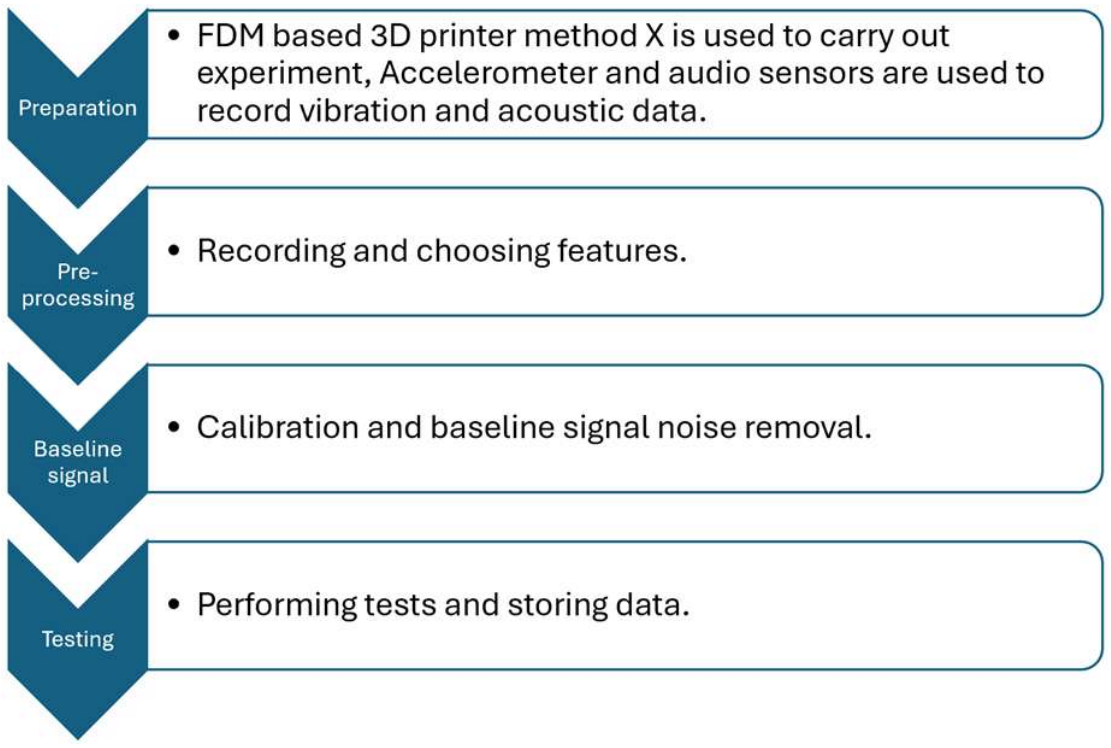

Fig. 2.1. Fundamental steps for in-situ monitoring and diagnostic for the FFF(Fused Filament Fabrication) process

In step 1, the FFF machine, Accelerometer sensor, Acoustic sensor, data acquisition (DAQ) device (Arduino uno), and laptop computer are used to set up the experiment system. Step 2 involves gathering the signals and choosing the features. In stages 3 and 4, data graphs are used to monitor and diagnose the machine state. The established setup comprises of the FFF Printer, Accelerometer, audio sensors, DAQ system, and a personal computer, as illustrated in Figure 3.1 The FFF equipment utilized in this investigation is Method X, a high-performance fiber composite FFF machine produced by MakerBot, (NYC, NY, USA). An accelerometer is attached to the extruder in a way that it can detect and measure horizontal movement in either direction. A DAQ unit is employed to process the signals which comprises of Arduino Uno. The digital signal processing module on the laptop is capable of processing real-time signals. The laptop used is XPS 9570 produced by Dell, and the signal processing software used is Arduino IDE. All data processing and testing is carried out in a MATLAB and Spectroid.

The FFF process specifications presented in this article are listed in Table 1.0. The material used is ABS-R, which provides a high-strength thermoplastic with good heat resistance, surface finish, and chemical resistance. The extrusion speed is 10mm/s.

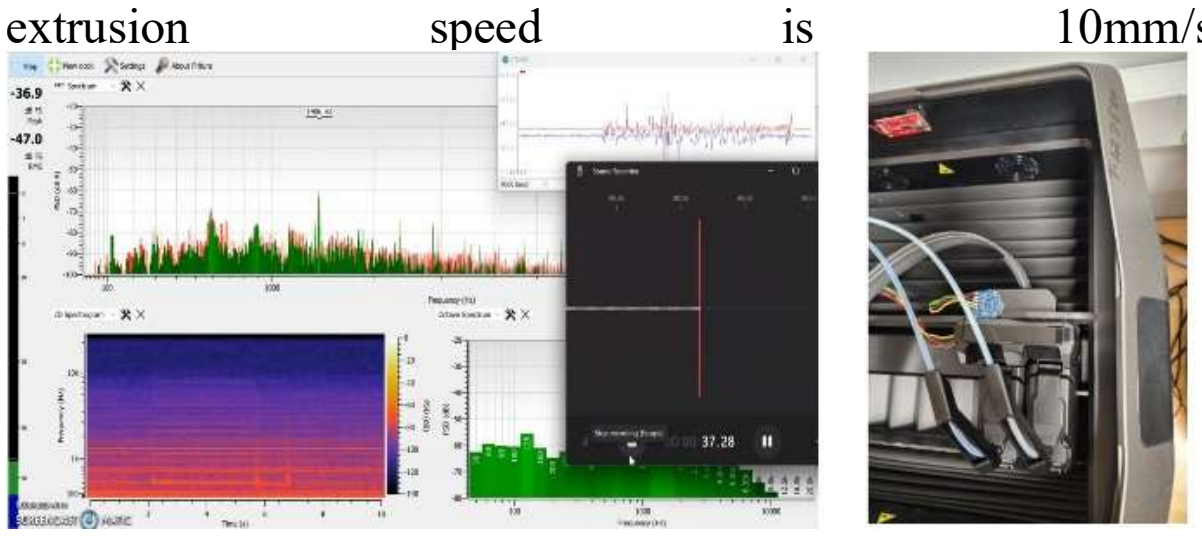

Fig 3.1. Audio sensor and accelerometer recording data simultaneously, with spectral visualization of audio signal and graphing of accelerometer data.

| DATA | VALUE |
|---|---|
| EXTRUDER TEMPERATURE | 270°C |
| MATERIAL TYPE | ABS – R |
| LAYER THICKNESS | 0.2mm |
| NOZZLE DIAMETER | 0.4mm |
| PRINT TYPE | 2 INCH SQUARE |

Table 1.0. Parameters of printer for the experimental setup.

Typical extrusion is achieved by maintaining the nozzle temperature within the designated temperature range for PLA during 3D printing. The obstructed and partially obstructed states were created by reducing the nozzle temperature. The runout state pertains to the excess material that flows out of the nozzle after extrusion has ceased. The loading state pertains to the insertion of the filament into the 3D printer before extrusion. Liu et al. [5] identified five machine states during their research on the acoustic emission monitoring of FDM machine operation, which are listed in Table 2.0. The normal state, as indicated, pertains to the continuous extrusion of the filament. The semi-blocked state refers to abnormal extrusion, where incomplete material is extruded through the nozzle. The blocked state arises when the nozzle is clogged and no material is extruded. The runout state pertains to the flow of molten material through the nozzle after 3D printing has ceased. Loading refers to the filament insertion before the commencement of a print.

| | |
|---|---|
| Normal | Extrusion speed 10 mm/s, nozzle temperature 210–220 °C |
| Semi-blocked | Extrusion speed 10 mm/s, nozzle temperature 170–180 °C |
| Blocked | Extrusion speed 10 mm/s, nozzle temperature 160 °C |
| Runout | Allowing the material to completely flow out after normal extrusion |
| Loading | Filament loading in the 3D printer before extrusion |

Table 2.0. Summary of the experimental conditions of Liu et al.

In another experiment, a mini audio sensor produced by sparkfun electronics was utilized as the audio sensor, with a frequency response range of 50 Hz to 16 kHz. A 40 dB gain was chosen for the amplifier, and hot glue was used to attach the sensor to the 3D printer's extruder body being studied. According to initial investigations, the signal frequencies were below 1 kHz, and hence, a sampling rate of 5 kHz was considered sufficient. The obtained signals were filtered using a band-pass filter from 1 to 2499 Hz. These positions were identified as the most convenient mounting points for placing the sensor close to the nozzles, taking into account the sensor's size.

# IV. OBSERVATIONS

As a baseline signal, only operating Y motor at a constant speed, The Y Axis stepper motor under load conditions generate Fundamental frequency @ 381 Hz and harmonics at multiple frequencies. See Fig 4.1.

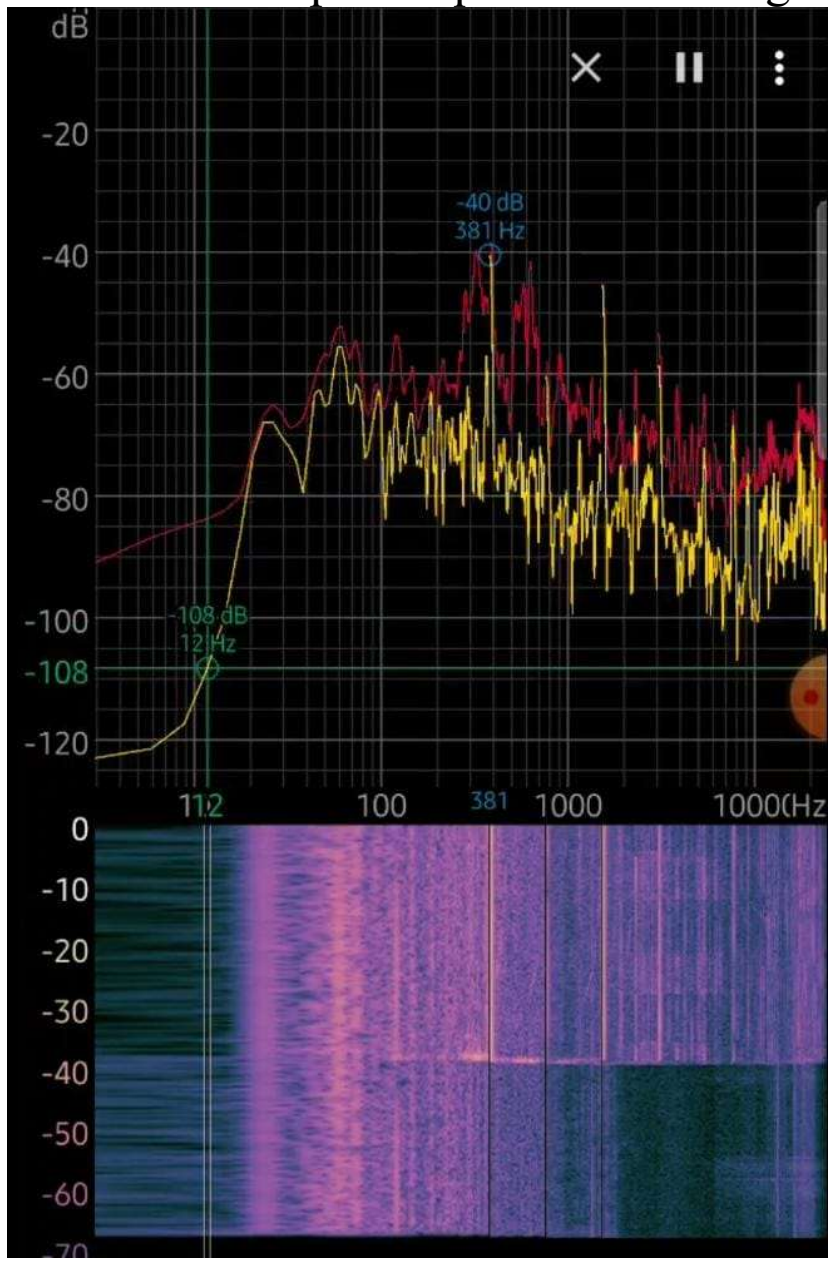

Fig. 4.1. A snapshot from Spectroid shows frequency visualization for Y motor.

During continuous printing the recorded frequency of audio signal changes abruptly because of the print head movement, this can be seen from the spectrogram below in Fig 4.2. The accelerometer readings shown in Fig 4.3. indicate that the vibration at extruder is at its peak during zigzag motion of normal operation.

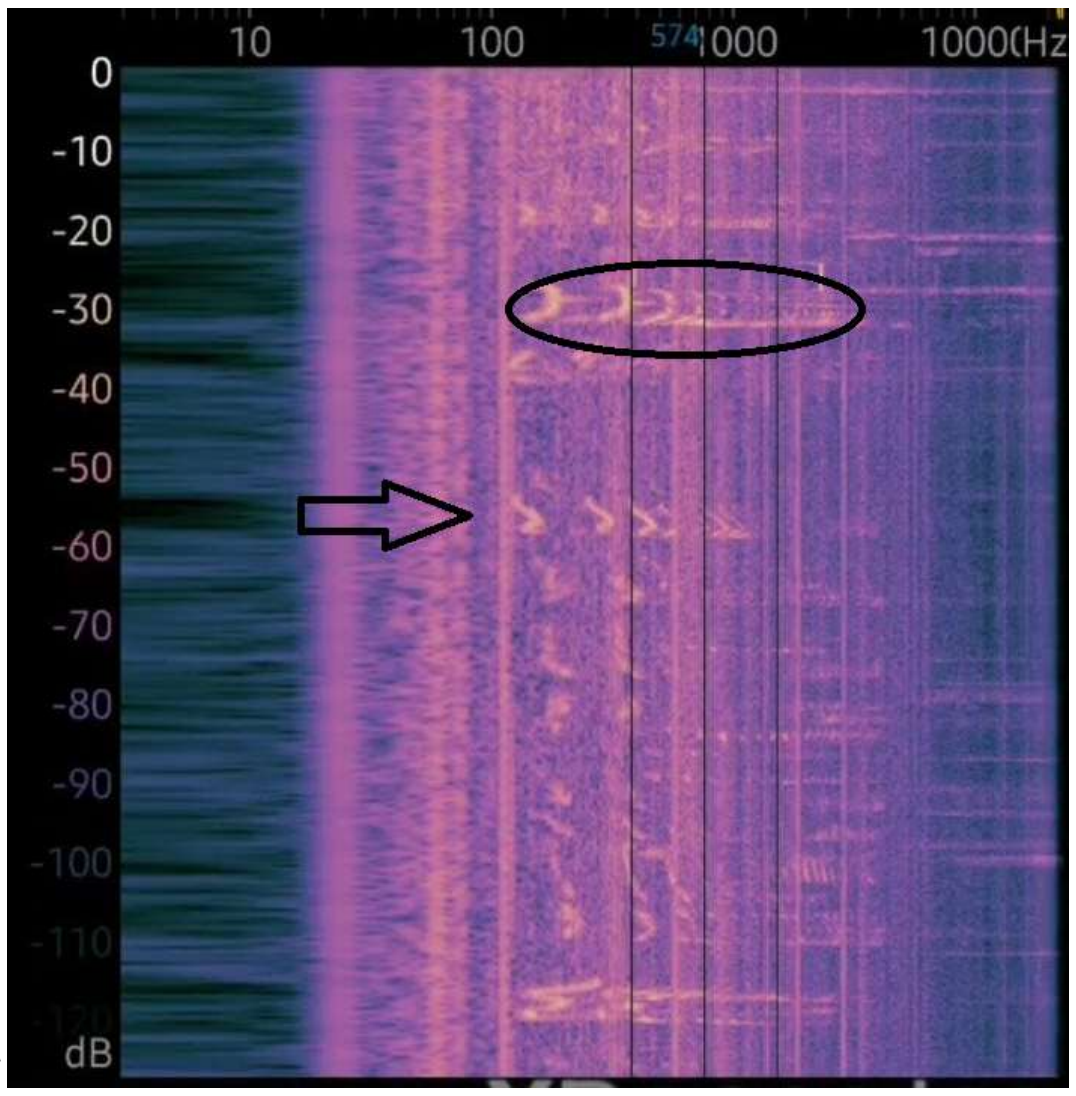

Fig. 4.2. Spectrogram shows frequency change between 100 and 1000 Hz.

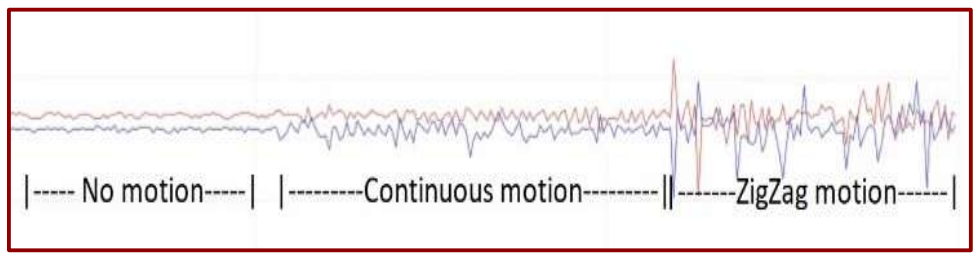

Fig. 4.3. The Accelerometer graph shows extruder movement patterns.

The recorded frequency changes during continuous printing can be attributed to the movement of the print head. In this scenario, high amplitude spikes in frequency were observed, ranging from 100 to 1000Hz. These spikes suggest the presence of multiple frequency components within the signal. The abrupt changes in frequency are likely a result of the dynamic nature of the printing process. As the print head moves, it introduces variations in the signal that manifest as spikes in the frequency domain.

The range of 100 to 1000Hz indicates a broad spectrum of frequencies, making it challenging to distinguish between the individual sounds originating from different parts of the printer machinery. The complexity of the signal, with its multiple frequency components, implies that various mechanical components within the printer contribute to the overall acoustic profile. Each component, such as gears, motors, or other moving parts, may generate distinct frequency signatures. The overlap of these frequencies during continuous printing creates a composite signal that is difficult to deconvolve.

To better understand and analyze the acoustic characteristics of the printer machinery, it may be necessary to employ advanced signal processing techniques. Additionally, monitoring the signal over time and correlating frequency changes with specific printer actions or movements may provide insights into the source of each frequency component.

The identified phenomenon persisted when an audio sensor was attached to the extruder body. This implies that the overlapping frequencies, ranging from 100 to 1000Hz, continue to pose challenges in isolating and focusing on specific sounds associated with the extrusion process. The extruder, being a critical component of a 3D printer responsible for material deposition, contributes its own set of acoustic signals, which, when mixed with other printer machinery sounds, creates a complex audio profile.

In such scenarios where distinguishing between individual sounds becomes intricate due to overlapping frequencies, the application of machine learning algorithms emerges as a promising solution[21]. These algorithms have shown efficacy in correlating frequency changes with specific actions or events, even in noisy environments[22]. By leveraging machine learning, it may be possible to train models that can identify and isolate the extrusion sound amidst the cacophony of other printer-related noises.

There is precedent for the application of deep learning algorithms in fault detection using acoustic signals. For instance, in a study involving an air compressor [20], deep learning techniques were employed to detect faults based on acoustic signals. This suggests that similar methodologies could be adapted for 3D printer fault detection.

## V. Conclusions

In this study, an analysis of accelerometer and audio sensor data was conducted on a Fused Deposition Modeling (FDM) based 3D printer. The findings indicate that both acoustic and vibration data carry valuable information for detecting faults. Notably, the vibration reaches its peak during the printer's zigzag motion, which could potentially lead to mechanical faults in the long term. Monitoring vibration data proves to be an effective method for early detection of faults.

The acoustic signatures produced by the 3D printer exhibit a complex overlap of various frequency patterns, making manual analysis challenging. However, there is significant potential in leveraging machine learning techniques to analyze these frequency patterns for fault detection purposes. Applying machine learning algorithms can enhance the efficiency of fault detection by automatically identifying and interpreting patterns within the acoustic data, thereby providing a more advanced and reliable method for identifying potential issues in the 3D printing process.

## Acknowledgment

This material is based upon work supported by the U.S. Department of Energy, National Nuclear Security Agency under Award Number DE-NA0003981. We express our sincere gratitude to the SPADAL Lab for granting us the valuable opportunity to conduct our research work. Additionally, we extend our appreciation to Honeywell and ASTERIX for their generous funding support. Their contributions have played a crucial role in making our research endeavors possible, and we are truly thankful for their commitment to advancing knowledge and innovation.